\documentclass[twocolumn,prb,showpacs,multicol,amsmath,amssymb]{revtex4}
\usepackage[dvips]{graphicx}
\usepackage{graphicx}
\usepackage{dcolumn}
\usepackage{bm}
\usepackage{graphics}
\usepackage{epsfig,color}
\usepackage[normalem]{ulem} 
\usepackage{soul}

\newcommand{\be}{\begin{equation}}
\newcommand{\ee}{\end{equation}}
\newcommand{\bea}{\begin{eqnarray}}
\newcommand{\eea}{\end{eqnarray}}

\begin{document}
\title{ Dynamics of coherence: Maximal quantum Fisher information vs. Loschmidt echo}
\author{Hadi Cheraghi}
\email[]{h.cheraghi1986@gmail.com}
\author{Saeed Mahdavifar}
\affiliation{ Department of Physics, University of Guilan, 41335-1914, Rasht, Iran}
\date{\today}
\begin{abstract}
We consider the dynamics of maximal quantum Fisher information (MQFI) after sudden quenches for the one-dimensional transverse-field Ising model. Our  results show,  the same as  Loschmidt echo, there is a  universality for the revival times i.e., they do not depend on the initial state and the size of the quench and are given by integer multiples of   $T_{rev} \simeq \frac{N}{2v_{max }}$, where $N$ is the system size and $v_{max }$ is the maximal group velocity of quasiparticles. Critically enhanced and decreased at revival and decay times as $T_{rev} \equiv T_{dec} $ are characterized by quenching from the order and disorder phases into the quantum phase transition respectively,  that can be utilized  to detect the quantum critical point (QCP). In some quenches crossed from the QCP, nonanalytic behaviors appear at  some times  due to the turning of the local observable from one direction to another because of identifying the maximum value. We name this phenomenon \textit{the dynamical MQFI transitions}, occurring at the critical times $t_c$. Interestingly, although no Fisher zero exists in the dynamics of MQFI, the first critical time emerged from the dynamical quantum phase transition is equal to the first time whose the logarithm of MQFI is  minimum.
In addition, we unveil the long-time run of MQFI indicates a signature of a nonequilibrium quantum phase transition at the QCP.
We also discuss the probability of arising of macroscopic  superpositions in the nonequilibrium dynamics of the system.
\end{abstract}
\maketitle

\section{Introduction}
Quantum coherence is crucially viewed as a quantum resource for building quantum information devices to obtain processes classically inhibited. 
Quantum optical methods provide an important set of tools for the control and the manipulation of coherence  in the out-of-equilibrium quantum many-body systems [{\color{blue}\onlinecite{1,2,3,4}}], yielding to the feasibility of making a quantum computer. 
Quantum coherence arising from quantum superposition is the key resource in various quantum information protocols such as quantum thermodynamics [{\color{blue}\onlinecite{5,6}}], metrology [{\color{blue}\onlinecite{7}}], cryptography [{\color{blue}\onlinecite{8}}], resource theory [{\color{blue}\onlinecite{9,10}}], and also exciton and electron transport in biomolecular networks [{\color{blue}\onlinecite{11}}]. 

Identifying key features of quantum coherence is highly controversial. There are some functions to qualify quantum coherence in the range of applicabilities such as trace distance [{\color{blue}\onlinecite{12}}], relative entropy [{\color{blue}\onlinecite{13}}], quantum correlation [{\color{blue}\onlinecite{14}}], the skew information [{\color{blue}\onlinecite{15}}], quantum Fisher information  [{\color{blue}\onlinecite{16}}], and Loschmidt echo (LE) [{\color{blue}\onlinecite{17}}].
Quantum Fisher information  is basically used to calculate the phase sensitivity that systems can provide in the imperfection of quantum measurement devices. In other words, it is applied for unknown parameters of a system by a quantum Cr\'amer-Rao bound which gives the achievable minimum estimation error [{\color{blue}\onlinecite{16,18}}]. As a witness of multipartite entanglement, it is demonstrated it characterizes topological states [{\color{blue}\onlinecite{181}}], and non-Gaussian many-body entangled states [{\color{blue}\onlinecite{182}}].
Quantum Fisher information  can be extensively probed in some context such as the relationship between
quantum coherence and quantum phase transition [{\color{blue}\onlinecite{19,20}}], quantum metrology [{\color{blue}\onlinecite{21}}], and quantum speedup limit time [{\color{blue}\onlinecite{22}}]. On the other hand, a standard way to study decoherence, stability, and complexity in dynamic processes is the LE. It  measures  the amount of the coherence spread between the initial and time-evolved states [{\color{blue}\onlinecite{{17}}}]. The LE has been explored
in several  problems including quantum chaos [{\color{blue}\onlinecite{23}}], equilibrium quantum phase transitions [{\color{blue}\onlinecite{24}}], work statistics [{\color{blue}\onlinecite{25}}], and non-markovianity [{\color{blue}\onlinecite{26}}].

 While the equilibrium phase transitions are perfectly well understood by means of standard methods like the mean-field theory and the renormalization group, the perceiving of the nonequilibrium dynamics is still vague. Inquiring on the context of the \textit{quantum quench}  has opened a wide window to study the out-of-equilibrium quantum systems [{\color{blue}\onlinecite{1,2,27,28,29}}]. Quantum quenches are done adiabatically slow [{\color{blue}\onlinecite{30}}] or  abruptly fast [{\color{blue}\onlinecite{31,32}}].
They are used  to probing ground-state phase transitions [{\color{blue}\onlinecite{321}}], detecting dynamical topology of entanglement-spectrum [{\color{blue}\onlinecite{322}}], localization and thermalization [{\color{blue}\onlinecite{323,324}}], and universality far-from-equilibrium [{\color{blue}\onlinecite{325,326}}]. 
Depending on the implementations, they can be studied by different methods  in many-body systems such as the Kibble-Zurek mechanism [{\color{blue}\onlinecite{33}}], measurement quench [{\color{blue}\onlinecite{34}}], and the dynamical quantum phase transition (DQPT) [{\color{blue}\onlinecite{35,36,37}}].

New insights can be gained on the fundamental question about the dynamics of the functions which relate to the measure of coherence in the systems. In this way, using two functions, the quantum Fisher information  and the LE, we are going to study nonequilibrium dynamics of coherence in the one-dimensional transverse-field Ising model, and seek what relationship between these two functions can be. To address this issue, we apply two conditions throughout our study those implemented in the DQPT: (i) the ground state of the initial Hamiltonian will be  chosen as the initial state of the system, (ii) the system will be examined at zero temperature. 
On one hand, using the two measures presented to the calculation of the macroscopic quantumness, we obtain an exact relationship for maximal quantum Fisher information (MQFI) and its dynamics. On the other hand, we compare the dynamics of the MQFI to the LE.
Our exact outcomes indicate there is an exciting relationship between the LE and the MQFI. That is, the revival times accept a promised universality as the initial state and the size of the quench are unimportant. In addition,  by  quenching  from the ferromagnetic (FM) and  paramagnetic (PM)  phases  into the quantum critical point (QCP) there will appear the revival and decay times  as $T_{rev} \equiv T_{dec} $ respectively.
Noteworthily, by crossing a quench from the QCP, two interesting results will obtain: (i) for some quenches, some non-analytical points as cusps emerge since of changing the direction of the local observable which detects MQFI. We call this event \textit{the dynamical MQFI transition}. (ii) Based on the definition of DQPT that is a logarithmic function of the LE,  we also consider the logarithm of the MQFI to examine its dynamics respect to the DQPT. The results show the first time in which the logarithm of MQFI is minimum  is exactly equal to the first critical time arising from DQPT. Additionally, we illustrate the long-time run of the MQFI is able to detect a nonequilibrium quantum phase transition at the QCP.

The rest of the paper goes as follows: In Sec. II, we present the model and review its exact solution. Section III is dedicated to
an analysis of the LE and the DQPT of the model.  In Sec. IV, the exact relationships for the MQFI and its dynamics are calculated.
Results and discussions are put in Sec. V. Here, in this section, we give detailed results for quenches into and crossed from the QCP. Further, we discuss the long-time run of the dynamical behavior of the MQFI to reveal nonequilibrium quantum phase transition and to investigate quenching within the same phase.

\section{Ising model in transverse field}

The Hamiltonian of 1D spin-1/2  Ising model in the presence of a transverse field  is given by
\begin{eqnarray}\label{eq1}
{\cal H} =  - \lambda \sum\limits_{n = 1}^N {\sigma _n^x\sigma _{n + 1}^x}  - \sum\limits_{n = 1}^N {\sigma _n^z}, 
\end{eqnarray}
where $\sigma _n^\mu $ is the $\mu$th Pauli matrix ($\mu=x,y,z$) at site $n$ and $\lambda>0$ denotes the power of  the ferromagnetic exchange coupling. 
The Hamiltonian has three symmetries, the translation invariance symmetry, the spin reflection symmetry, and global
phase flip symmetry as $[U,{\cal H}] = 0$ where $U = \prod\nolimits_{n = 1}^N {\sigma _n^z} $ [{\color{blue}\onlinecite{38}}].
The Hamiltonian conserves the parity of the particle number and acts differently on the even (Neveu-Schwarz) and odd (Ramond) subspaces. In the fermionic Fock space, the Hamiltonian in the two subspaces is formally the same if one imposes antiperiodic boundary condition for the even and periodic boundary condition for the odd subspace that in wave-number space these boundary conditions translate to different quantization as momentum quantization  in half-integer and in integer multiples of $\frac{2\pi}{N}$ respectively [{\color{blue}\onlinecite{38,39}}].
In the thermodynamic limit, the ground states of the odd and even subspaces become degenerate and one recovers the two fully polarized ferromagnetic ground states. Here, the  periodic boundary condition $\sigma _{n+1}^\mu=\sigma _1^\mu $ is considered. The model exhibits a quantum phase transition at  $  \lambda_c = 1$ from  a FM phase ( $\lambda > 1$) to a PM phase ($\lambda < 1$). 
At $\lambda_c =1$ where  the symmetry break downs and the system undergoes a quantum phase transition,  superposition states in the  ground state of  the system manifest.
It is shown these superpositions behave as macroscopic superpositions that adhere to the scaling of the effective size [{\color{blue}\onlinecite{40}}]. 

The Hamiltonian  is  integrable and can be mapped to a system of free fermions and therefore be solved exactly. By applying the Jordan-Wigner transformation [{\color{blue}\onlinecite{41}}], a Fourier transformation as ${a_n} = \frac{1}{{\sqrt N }}\sum\limits_k {{e^{ - ikn}}} {a_k}$ where $a_n$ is fermionic operator, and also Bogoliobov transformation as ${a_k} = \cos ({\theta _k}){\alpha _k} + i\sin ({\theta _k})\alpha _{ - k}^{\dag}$, the quasi-particle diagonalized  Hamiltonian obtains as
\begin{equation}\label{eq2}
{\cal H} = \sum\limits_k {{\varepsilon }_k  [ {\alpha _k^\dag {\alpha _k} - 1/2} ]  },
\end{equation}
where the energy spectrum is $\varepsilon_k = \sqrt{{\cal A}_k ^2+{\cal B}_k ^2}$ with
\begin{eqnarray}\label{eq3}
{\cal A}_k=-2[\lambda\cos(k)+1] ~~~~~;~~~~~{\cal B}_k= 2 \lambda \sin(k),
\end{eqnarray}
and  $\tan (2{\theta _k}) =- \frac{{{{\cal B}}_k }}{{{{\cal A}}_k }}$.


\section{Loschmidet echo and Dynamical Quantum Phase Transition}

Manipulating a quantum system requires the knowledge of how it evolves with passing time. There are many ways to drive a physical system away from equilibrium. 
One simplest controllable plan is quench dynamics,  putting the system  in an equilibrium state described with the Hamiltonian $\mathcal{H}_1(\lambda_1) $ and a well-defined  initial state  $\left| {{\Psi _0}} \right\rangle $, afterward,   taking out-of-equilibrium by suddenly changing the control parameter from its initial value to its final value, $\lambda_1 \longrightarrow  \lambda_2$. The final Hamiltonian and its time-evolved state will be as  ${\cal H}_2(\lambda_2) $ and $\left| {\Psi (t)} \right\rangle  = {e^{ - it\mathcal{H}_2(\lambda_2)}}\left| {{\Psi _0}} \right\rangle $, respectively. 
Since $\left| {\Psi (t)} \right\rangle $ typically consists of many excited states of $\mathcal{H}_2(\lambda_2)$ with a non-thermal distribution, its time evolution provides a unique venue for investigating issues in nonequilibrium quantum statistical mechanics such as quantum decoherence  [{\color{blue}\onlinecite{{17}}}], and equilibrium quantum phase transitions [{\color{blue}\onlinecite{{42}}}]. Here, we examine our problem with two conditions: \\
\begin{list}{(i)}
\item Fixing the initial state of the system into the ground state of the initial Hamiltonian.
\end{list}
\begin{list}{(ii)}
\item Considering the system at zero temperature. \\
\end{list}
Using these characteristics, the LE can be defined as  $LE(t) = \prod\limits_{k } |{{{\cal L}_k}} (t)|$ where ${\cal L}_k(t) =  {\left\langle {{\Psi _0}} \right|{e^{ - i {\cal H}_2(\lambda_2) t}}\left| {{\Psi _0}} \right\rangle }$.
A simple calculation gives an exact expression for the LE
\begin{eqnarray}\label{eq4}
LE(t) =  \prod\limits_{k > 0} {\left| \cos ^2(\Phi _k) + \sin ^2(\Phi _k)e^{ - 2it\varepsilon _k^{(2)}} \right|},
\end{eqnarray}
where  ${\Phi _k} = \theta _k^{(2)}- \theta _k^{(1)}$ is the difference between the Bogoliubov angles diagonalizing the post-quench and pre-quench Hamiltonians, respectively. 
The DQPT is discovered with the formal similarity of the partition function $Z(\beta ) = tr\left( {{e^{ - \beta \mathcal{H}}}} \right)$  bounded by the initial state  $\left| {{\Psi _0}} \right\rangle$  as $Z(z) = \left\langle {{\Psi _0}} \right|{e^{ - z\mathcal{H}}}\left| {{\Psi _0}} \right\rangle$ with $z \in \mathbb{C}$ so that it represents the LE for $z =it$. Likewise, the DQPT is defined as the rate function of the return probability in the form [{\color{blue}\onlinecite{{35}}}]
\begin{eqnarray}\label{eq5}
r_{LE}(t)=  -  \frac{1}{N}\log |LE(t)|^2,
\end{eqnarray}
where $N$ is the system size that should be large enough. For quenches crossed from  a QCP,  this quantity exhibits non-analytic behaviors in the form of cusps  taken place in Fisher zeroes [{\color{blue}\onlinecite{43}}]  that appear periodically at the critical times 
\begin{equation}\label{eq6}
t_n^* = t^*(n + \frac{1}{2}),~~~n = 0,\pm 1,\pm 2,...
\end{equation}
with  ${t^*} = \frac{\pi }{{\varepsilon _{{k^*}}^{(2)}}}$ where $k^*$ is the particular mode  driven from $\cos(2 \Phi _{k^*}) = 0$ as  $\cos ({k^*}) =  - \frac{{1 + {\lambda _1}{\lambda _2}}}{{{\lambda _1} + {\lambda _2}}}$  which leads to vanish the argument in the logarithm in ({\color{blue}{\ref{eq5}}}). It should note some models disclose two Fisher zeroes [{\color{blue}\onlinecite{{44}}}] although some others nothing [{\color{blue}\onlinecite{{45}}}].


\section{Dynamics of maximal quantum Fisher information}


\subsection{Maximal Quantum Fisher Information}

Using the quantum quench and two applicable formulations to study the macroscopic superposition in a many-body quantum system, we  obtain a relationship for calculating the dynamical behavior of the MQFI after sudden quenches. The  two measurable quantities that quantify the degree of the macroscopic quantumness of a given state are one based on the correlation of local observables on many sites [{\color{blue}\onlinecite{{46}}}] and another based on the quantum Fisher information  [{\color{blue}\onlinecite{{47}}}].
Let ${\cal A}$ be the set of all additive operators $A$ as  $A = \sum\nolimits_{n = 1}^N {{A_n}} $.
The former measurement defined as the variance on the additive operator  $A $  is given by
\begin{eqnarray}\label{eq7}
{{\cal V}_\Psi }(A): = \mathop {\max }\limits_{A \in {\cal A}}  \left[ \left\langle \Psi  \right|{{A}^2}\left| \Psi  \right\rangle  - \left\langle \Psi  \right|{ A}{\left| \Psi  \right\rangle ^2}   \right],
\end{eqnarray}
where the maximum is taken over all Hermitian additive operators. Every operator $A_n$ acts nontrivially on the $n$th particle and $||A_n||=1$.
A straightforward calculation exhibits any product state $\left| \Psi  \right\rangle  = {\left| \Phi  \right\rangle ^{ \otimes N}}$ is proportional to $N$ as ${{\cal V}_\Psi }({ A}) = N{{\cal V}_\Phi }({ A})$, and a Greenberger-Horne-Zeilinger state [{\color{blue}\onlinecite{{48}}}]  $\left| {GHZ} \right\rangle  = \frac{1}{{\sqrt 2 }}({\left|  \uparrow  \right\rangle ^{ \otimes N}} + {\left|  \downarrow  \right\rangle ^{ \otimes N}})$  which is a macroscopic quantum superposition is proportional to ${{\cal V}_{GHZ}}({M_z}) = {N^2}$. Based on these results, it is defined $p\textendash$index of a pure state  $\left| \Psi  \right\rangle$ as 
\begin{eqnarray}\label{eq8}
\mathop {\max }\limits_{A \in {\cal A}} {{\cal V}_\Psi }({ A}) = {\cal O}({N^p}),~~~~N~large
\end{eqnarray}
A fully product state gives $p=1$. On the other hand, $p=2$ contains a superposition of macroscopically distinct
states, in the sense of nonvanishing  the relative fluctuation  in the thermodynamic limit, $N \longrightarrow \infty$. As a consequence, one can find  $p>1$ is an entanglement witness for pure states. To find the additive operator $ A$, we should start from the general form of it as $A_n = \overrightarrow \sigma _n.\overrightarrow n $ where $\overrightarrow n  = (\sin \beta \cos \phi ,\sin \beta \sin \phi ,\cos \beta )$. Using ({\color{blue}\ref{eq7}}) we obtain
\begin{eqnarray}\label{eq9}
{{\cal V}_\Psi }({A}) &=& {\sin ^2}(\beta )\left[ {\left\langle {{X^2}} \right\rangle {{\cos }^2}(\phi ) + \left\langle {{Y^2}} \right\rangle {{\sin }^2}(\phi )} \right] \nonumber\\
 &+& {\cos ^2}(\beta )\left[ {\left\langle {{Z^2}} \right\rangle  - {{\left\langle Z \right\rangle }^2}} \right],
\end{eqnarray}
where $X = \sum\limits_{n = 1}^N {\sigma _n^x} $  with similar definitions for $Y$ and $Z$. In ({\color{blue}\ref{eq9}}) we have used $\left\langle X \right\rangle  = \left\langle Y \right\rangle  = \left\langle XZ\right\rangle  =\left\langle YZ \right\rangle=0$
and the fact that  $\left\langle XY+YX \right\rangle$ vanishes due to the reality of the Hamiltonian and Hermiticity of the operator $XY+YX$.

\begin{figure}[t]
\centerline{\includegraphics[width=0.8\linewidth]{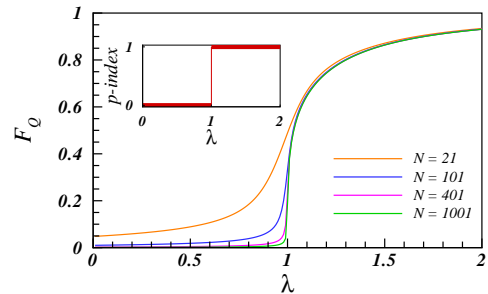}}
\caption{(color online). The MQFI versus $ \lambda$ at $t=0$ measured by  ${{\cal F}_{\cal Q}} = [{\mathop {\max }\limits_{A \in {\cal A}} {\cal F}(\rho ,A)}]/{N^2}$  for  sizes $N=21, 101, 401, 1001$. The inset shows that the $p\textendash$index at thermodynamics limit, $N \to \infty $, discloses a jump at $ \lambda_c=1$ where the quantum phase transition takes place.}
\label{Fig1}
\end{figure}

\begin{figure*}[t]
\centerline{
\includegraphics[width=0.35\linewidth]{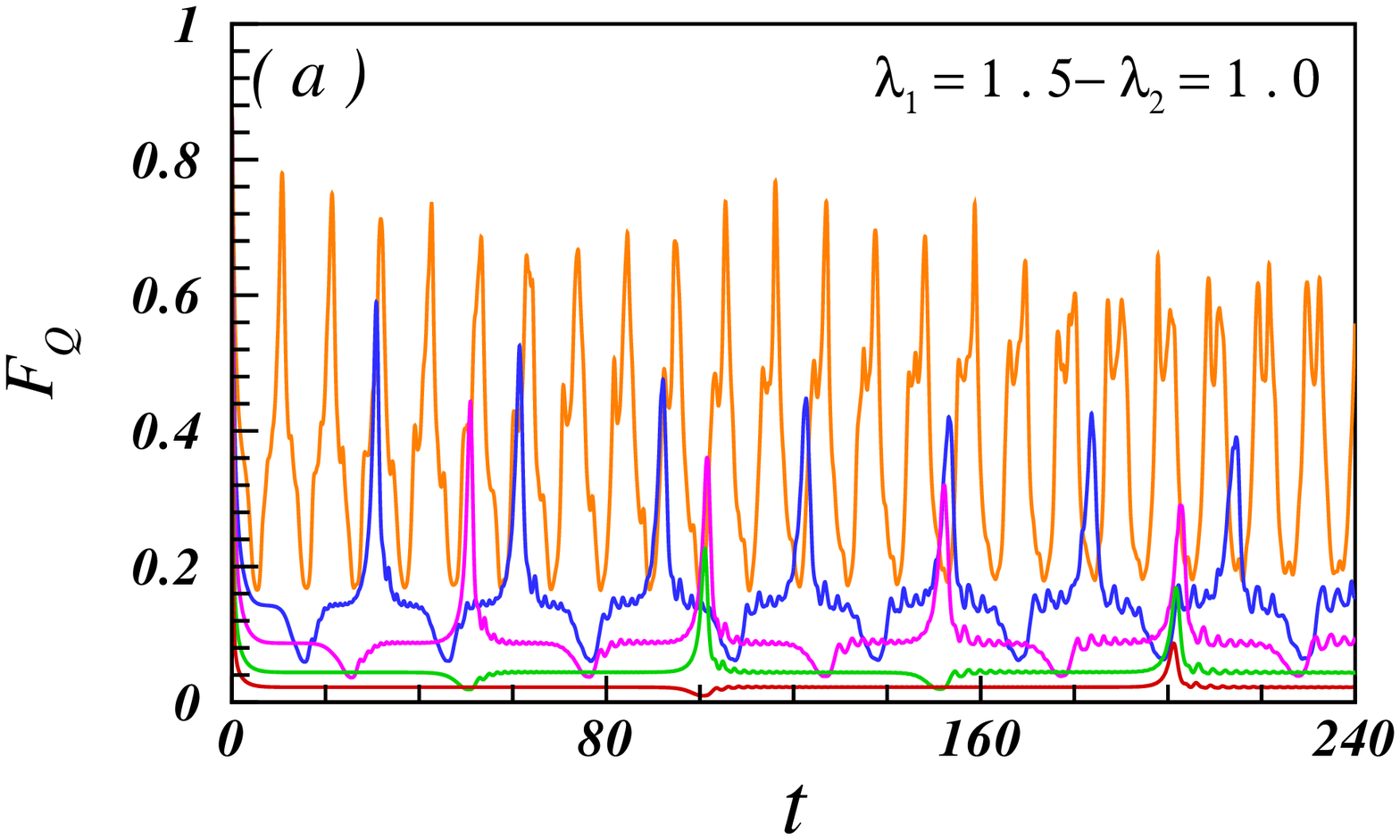}
\includegraphics[width=0.35\linewidth]{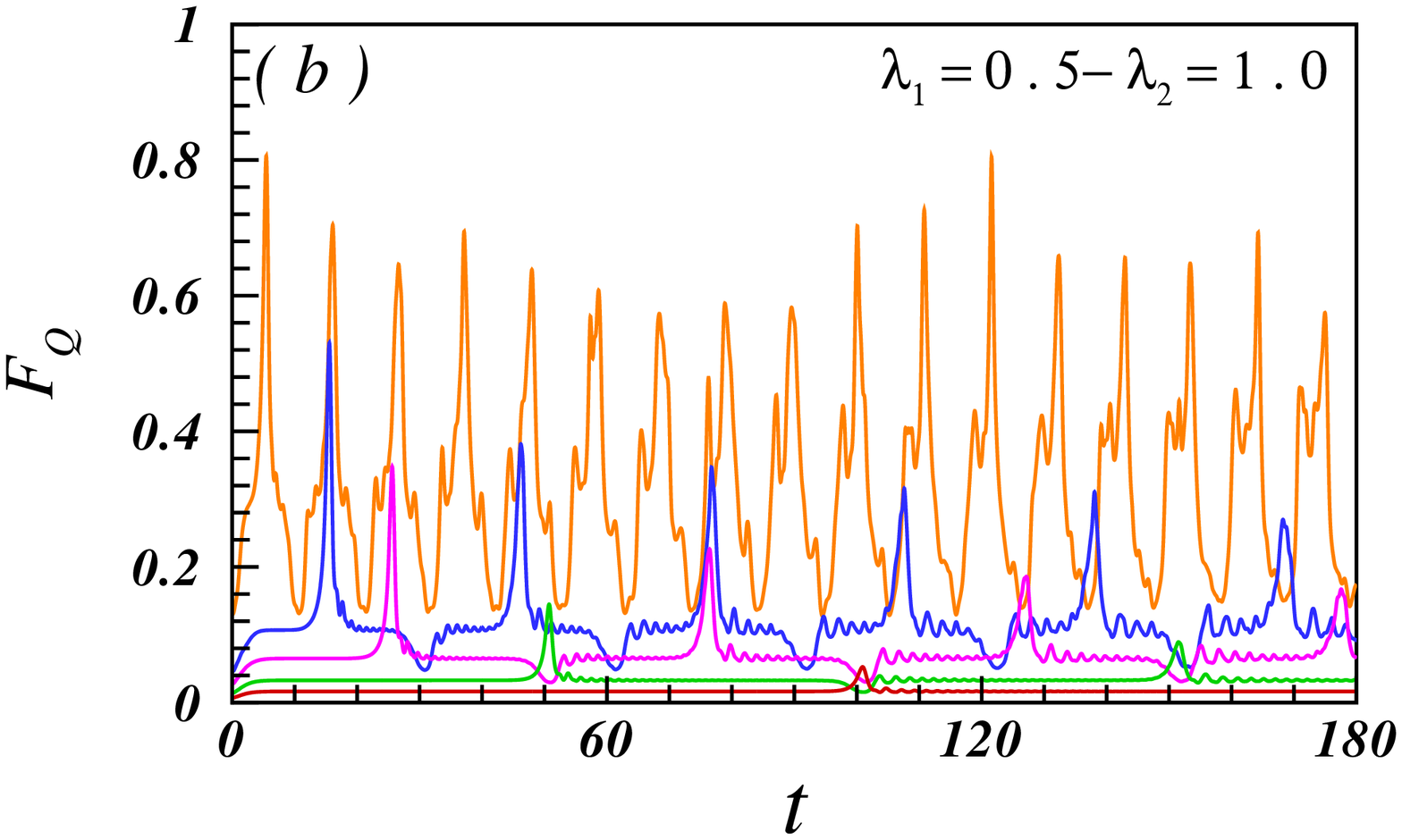}
\includegraphics[width=0.187\linewidth]{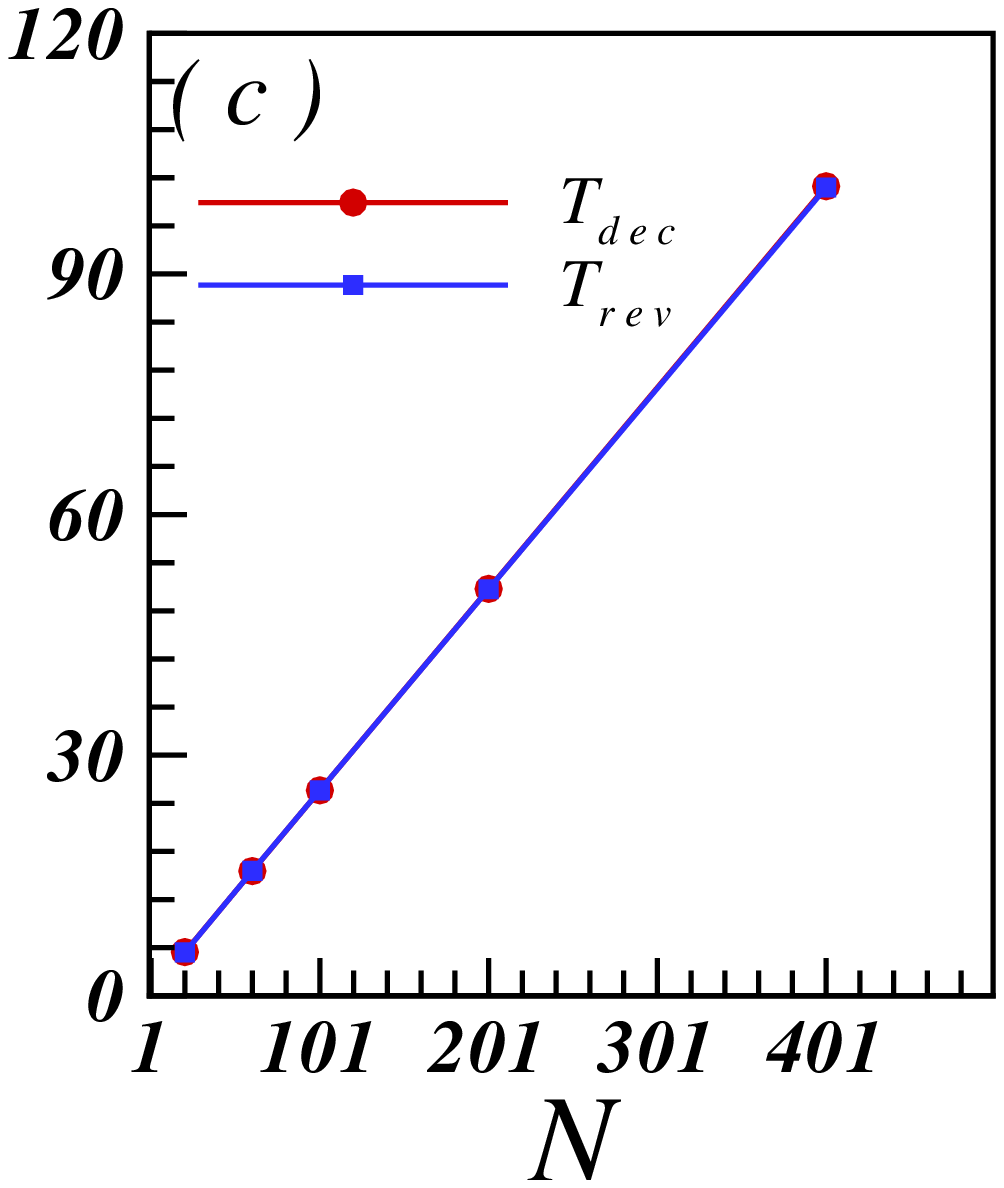}}
\caption{(color online). Dynamics of MQFI  measured by ${\cal F}_{\cal Q}$ for quenches from (a) $ \lambda_1=1.5$, and (b) $ \lambda_1=0.5$, into the QCP $ \lambda_2=\lambda_c=1.0$ for different sizes $N=21,61,101,201,401$ (from orange to red) respectively. (c) Scaling of the first revival $T_{rev}$ and decay $T_{dec}$  times  respect to the different system sizes for the quenches corresponding to (a \& b), revealing $T_{{rev}/{dec}}  \propto N$.}
\label{Fig2}
\end{figure*}

Another measure of macroscopicity was defined according to the quantum Fisher information. In the point of view of a quantum state, this is a measure of how fast given state changes under a given evolution that originally was disclosed in the context of phase estimation [{\color{blue}\onlinecite{{49}}}].  It is used as an apparatus to indicate how useful a quantum state is for quantum metrology [{\color{blue}\onlinecite{{50,51}}}], and to provide a lower bound on multipartite entanglement [{\color{blue}\onlinecite{{52,53}}}].
 Additionally, it is demonstrated the macroscopic quantumness, $N_{eff}$, of a many-body state is related to its MQFI with respect to all extensive observables $ A$ [{\color{blue}\onlinecite{{47}}}]. For a general initial quantum state $\rho $ of $N$ particles, this measure is defined in the form
\begin{eqnarray}\label{eq10}
{N_{eff}}: = \frac{1}{{4N}}\mathop {\max }\limits_{A \in {\cal A}} {\cal F}(\rho ,A),
\end{eqnarray}
where ${\cal F}(\rho ,A)$ is  quantum Fisher information.
It should be noted that the quantum Fisher information  reduces to 4 times the variance for pure states. For all pure states $\rho  = \left| \Psi  \right\rangle \left\langle \Psi  \right|$, the range of the effective size is $1 \le {N_{eff}} \le N$. That is, macroscopic quantum behavior arises as  linear in the system size $N_{eff}={\cal O}(N)$ while product states gives  $N_{eff}={\cal O}(1)$. In other words, entangled states can exhibit much larger Fisher information than separable states. Consequently, a combination of two mentioned measures  takes 
\begin{eqnarray}\label{eq11}
 \mathop {\max }\limits_{A \in {\cal A}}{\cal F}(\rho ,A)= \mathop {\max }\limits_{A \in {\cal A}} {{\cal V}_\Psi }( A),
\end{eqnarray}
To avoid dealing with large values, we define a parameter as
 ${\cal F}_Q=\frac{\mathop {\max }\limits_{A \in {\cal A}} {\cal F}(\rho ,A)}{N^2}$. Finally, the outcome is 
\begin{eqnarray}\label{eq12}
{\cal F}_Q &=&\frac{1}{N^2} \mathop {\max }\limits_{\beta, \phi}   \left\{ {\sin ^2}(\beta )\left[ {\left\langle {{X^2}} \right\rangle {{\cos }^2}(\phi ) + \left\langle {{Y^2}} \right\rangle {{\sin }^2}(\phi )} \right]  \right. \nonumber\\
 &+&\left. {\cos ^2}(\beta )\left[ {\left\langle {{Z^2}} \right\rangle  - {{\left\langle Z \right\rangle }^2}} \right]  \right\},
\end{eqnarray}
 with $0 \le {{\cal F}_{\cal Q}} \le 1$. The angles $\beta $ and $\phi $ change in intervals  $\beta=[0,\pi] $ and $\phi=[0,2\pi] $, and the process of maximizing must be done in these regions. The results show for the Hamiltonian ({\color{blue}\ref{eq1}}), at $t=0$ the maximum value occurs at $\overrightarrow n  = \overrightarrow x $ with $(\beta=\frac{\pi}{2},\phi=0)$. However, when the system evolves with passing time, we must consider the general form of ({\color{blue}\ref{eq12}}).
In the rest of the paper, we use the parameter ${\cal F}_Q$ to consider MQFI.

\begin{figure*}[t]
\centerline{
\includegraphics[width=0.35\linewidth]{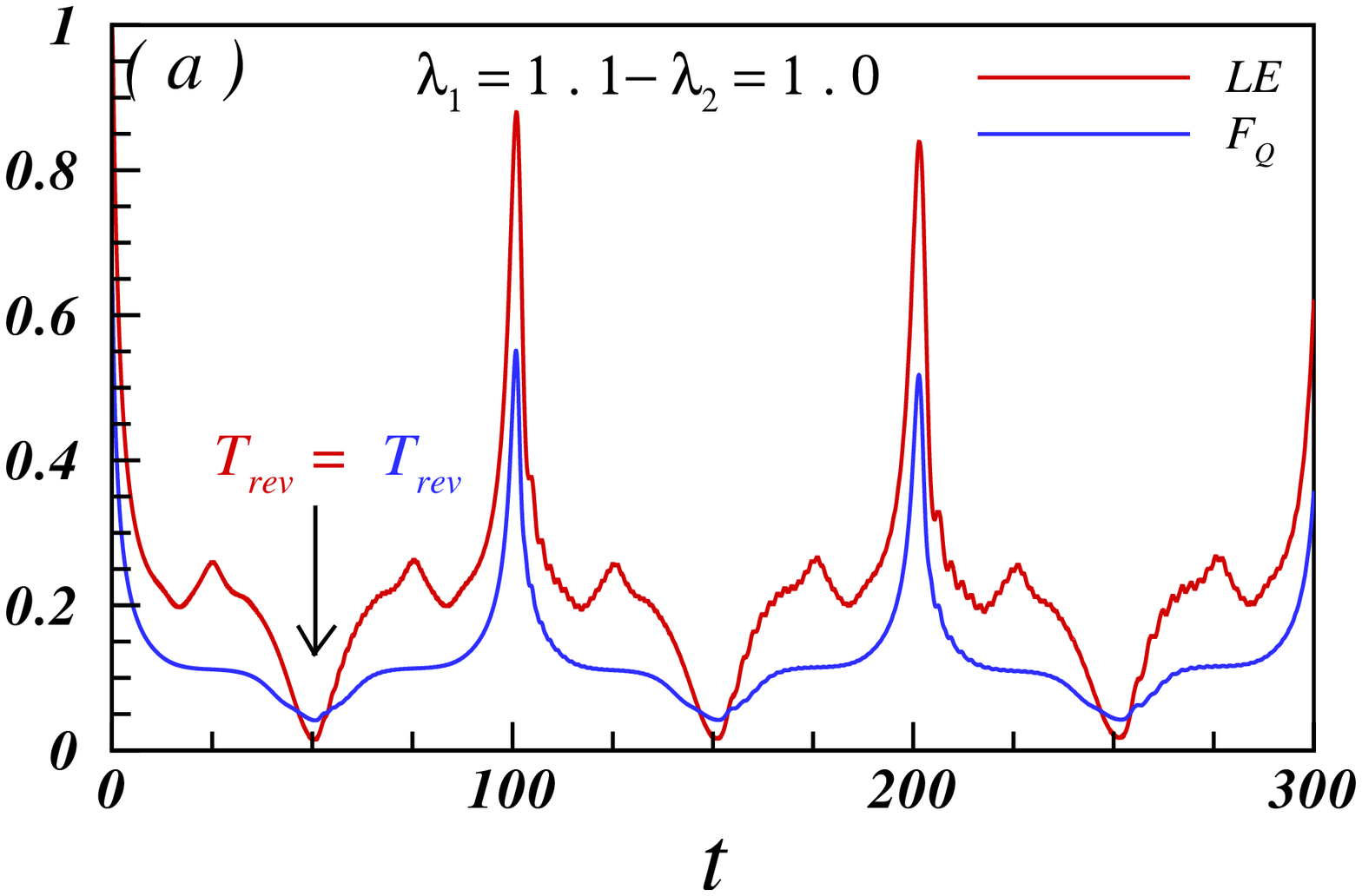}
\includegraphics[width=0.35\linewidth]{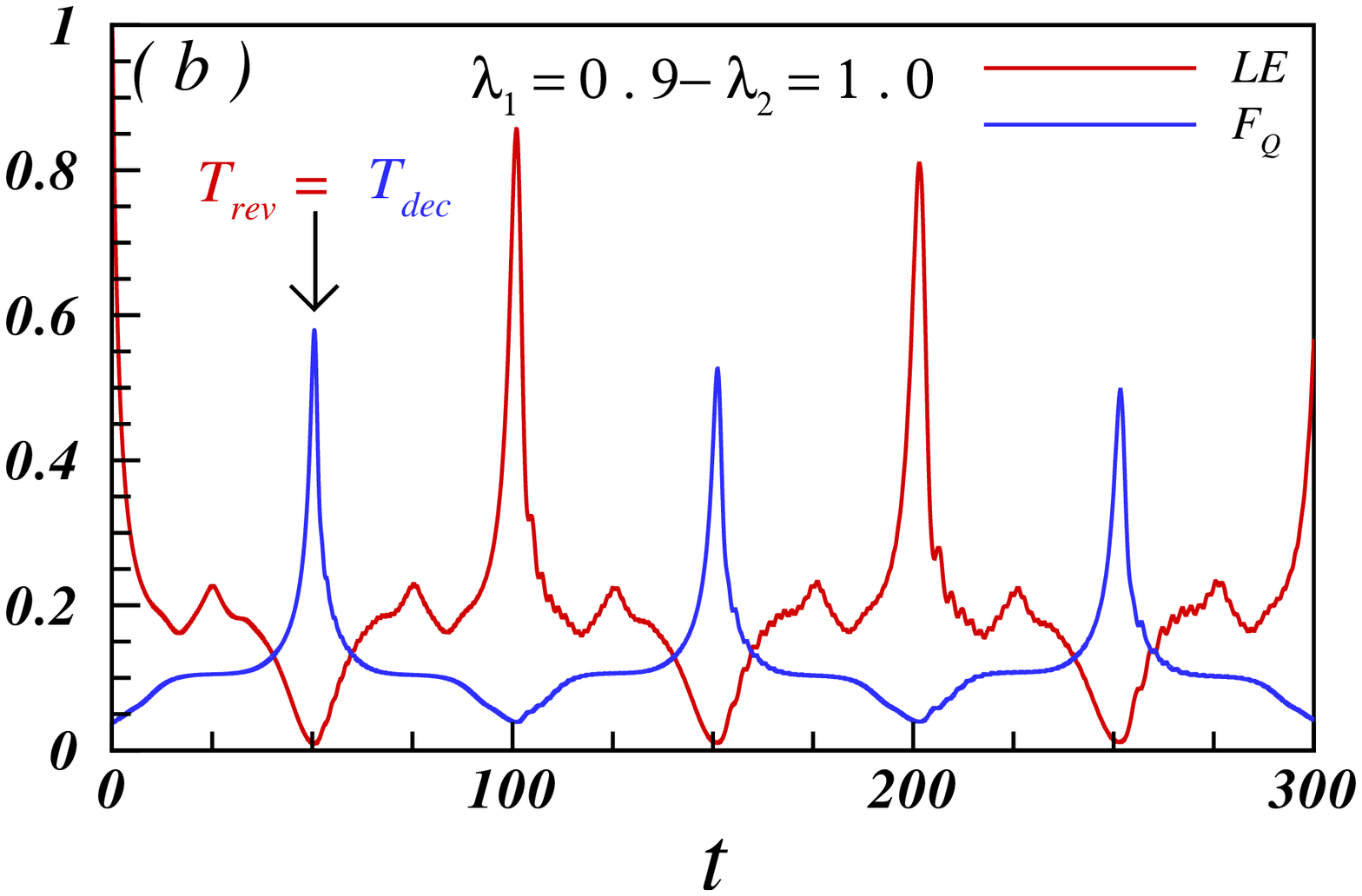}
\includegraphics[width=0.35\linewidth]{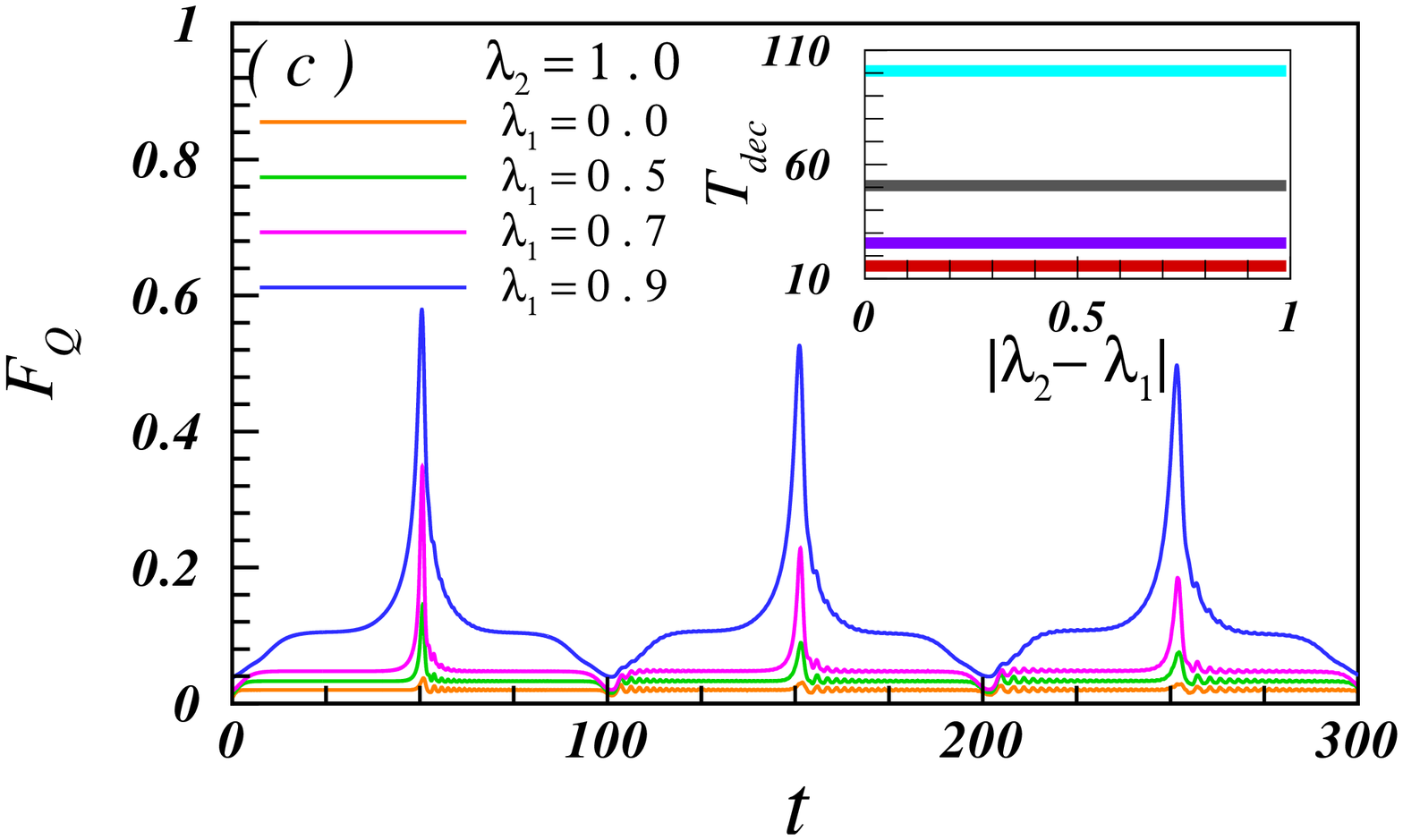}}
\caption{(color online). All plots are drawn for size $N=201$. Dynamics of MQFI  measured by  ${\cal F}_{\cal Q}$, and  LE for quenches as (a) $ \lambda_1=1.1$, and (b) $ \lambda_1=0.9$, into the QCP $ \lambda_2=\lambda_c=1.0$. As explicitly seen, ${\cal F}_{\cal Q}$ and LE are quite matching in the revival and decay times. (c) Quenches from $ \lambda_1=0.0, 0.5, 0.7, 0.9$ established from the PM phase into the QCP that displays the shape of structure of all quenches are the same. The insert in (c) is the first decay time $T_{dec}$ versus $ |\lambda_2-\lambda_1|$ for quenches from region $0 \le {\lambda _1} < 1$ into $ \lambda_2=1.0$ for sizes $N=61,101,201,401$ (from red to cyan). As clearly seen, for a given system size,  the first decay time stays on a constant value.}
\label{Fig3}
\end{figure*}


\subsection{Dynamical Behavior}

Based on analytical insights, it is obvious $n=1$ gives  $\left\langle {\sigma _1^\mu (t)\sigma _1^\mu(t)} \right\rangle  = 1$ with $\mu=x,y,z$. On the other hand, one can write
\begin{eqnarray}\label{eq13}
\left\langle {{\eta ^2}} (t)\right\rangle  = N\left[ {1 + \sum\limits_{n = 1}^{N - 1} {G_n^{\mu \mu }}(t) } \right]~~;~~\eta  = X,Y,Z
\end{eqnarray}
 where the two point functions are shown as the form $G ^{\mu\mu}_n(t): = \left\langle {\sigma _1^\mu(t)\sigma _{1+n}^\mu(t)} \right\rangle $.
Thus, the determination of the ${\cal F}_{\cal Q}(t)$  is reduced to calculation of the two point functions $G ^{\mu\mu}_n(t)$. 
 Since the nonlocal nature of the Jordan-Wigner transformation, the calculation of  the two-point spin functions are quite nontrivial.
Consequently, we make use of the well-known relations  to calculate  them determined in Refs. [{\color{blue}\onlinecite{{39,54}}}]   for $n \ge 1$ as 
\begin{eqnarray}\label{eq14}
G _n^{xx}(t)&:=& \left| {\begin{array}{*{20}{c}}
{{G_{ - 1}}(t)}&{{G_{ - 2}}(t)}&{...}&{{G_{ -n}}(t)}\\
{{G_0}(t)}&{{G_{ - 1}}(t)}&{...}&{{G_{ - n + 1}}(t)}\\
\begin{array}{l}
.\\
.\\
.
\end{array}&\begin{array}{l}
.\\
.\\
.
\end{array}&\begin{array}{l}
.\\
.\\
.
\end{array}&\begin{array}{l}
.\\
.\\
.
\end{array}\\
{{G_{n - 2}}(t)}&{{G_{n- 3}}(t)}&{...}&{{G_{ - 1}}(t)}
\end{array}} \right|,
\nonumber\\
G_n^{yy}(t) &:=& \left| {\begin{array}{*{20}{c}}
{{G_1(t)}}&{{G_0(t)}}&{...}&{{G_{2 - n}(t)}}\\
{{G_2}(t)}&{{G_1}(t)}&{...}&{{G_{3 - n}(t)}}\\
{\begin{array}{*{20}{c}}
.\\
.\\
.
\end{array}}&{\begin{array}{*{20}{c}}
.\\
.\\
.
\end{array}}&{\begin{array}{*{20}{c}}
.\\
.\\
.
\end{array}}&{\begin{array}{*{20}{c}}
.\\
.\\
.
\end{array}}\\
{{G_n}(t)}&{{G_{n - 1}(t)}}&{...}&{{G_1}(t)}
\end{array}} \right|,
\nonumber\\
G_n^{zz}(t) &:=& G_0^2(t) - {G_n}(t){G_{ - n}}(t).
\end{eqnarray}
and $\left\langle Z \right\rangle  = NG_0(t)$  where
\begin{eqnarray}\label{eq15}
G_n(t)&=&  - \frac{2}{N}\sum\limits_{k > 0} {\left\{ {\cos (2{\Phi _k})\cos \left[ kn+ 2\theta _k^{(2)} \right]} \right.} \nonumber\\
 &+& \left. {\sin (2{\Phi _k})\sin \left[kn+ 2\theta _k^{(2)} \right]\cos \left( {2\varepsilon _k^{(2)}t} \right)} \right\},\nonumber \\
\end{eqnarray}
that $k=\frac{2\pi m}{N}$ with $m=0,1,...,\frac{1}{2}(N-1)$. One can explicitly see at $t=0$, when no quench is done, ({\color{blue}\ref{eq15}}) converts to $G_n$ presented in Ref.~[{\color{blue}\onlinecite{{54}}}].


\section{RESULTS AND DISCUSSIONS}

In  Fig.~{\color{blue} \ref{Fig1}} we have depicted MQFI measured by ${\cal F}_{\cal Q}$ versus $\lambda$ at $t=0$,  for different system sizes $N=21,101,401,1001$. As one can see, when the system size is large enough at $\lambda=0$  where the spins are completely aligned in the z-direction, ${\cal F}_{\cal Q}$ is zero. This is because of the system settles at a complete disorder phase where entanglement vanishes [{\color{blue}\onlinecite{53}}]. 
As the value of $\lambda$ increases, the value of ${\cal F}_{\cal Q}$ increases only slightly. This increment will continue until $\lambda$ reaches to $\lambda_m(N)$. In $\lambda_m(N)$, there is a discontinuity in the ${\cal F}_{\cal Q}$ and the concavity of the graph changes. After that, by enhancing the value of $\lambda$, the value of ${\cal F}_{\cal Q}$ tends to the value of one. At the thermodynamic limit as $N$ approaches $\infty$,  we have $\lambda_m(\infty )=\lambda_c=1$, and for $\lambda<\lambda_c$, the value of ${\cal F}_{\cal Q}$ is almost zero. It is  illustrated for all sizes, the derivative of ${{\cal F}_{\cal Q}}$ is maximum at $\lambda_m(N)$ [{\color{blue}\onlinecite{40}}].
Moreover, the inserted figure indicates that at the thermodynamic limit the $p\textendash$index suddenly goes from 0 to 1, and a discontinuous transition manifests. Hence, for $\lambda>\lambda_c$, the ground state of the system is a macroscopic superposition. 
 Consequently, these characteristics imprint MQFI can correctly disclose the QCP if the system size will be large enough. 
 It is  shown  the effective size of macroscopic superposition between the two symmetry breaking states  grows to the scale of the system size as  $1 - {\lambda _m}(N) \sim {N^{ - 1.96}}$. Corresponding to ({\color{blue}\ref{eq10}}), this scaling can be interpreted as the scaling of MQFI near to the QCP.

\begin{figure*}[t]
\centerline{
\includegraphics[width=0.35\linewidth]{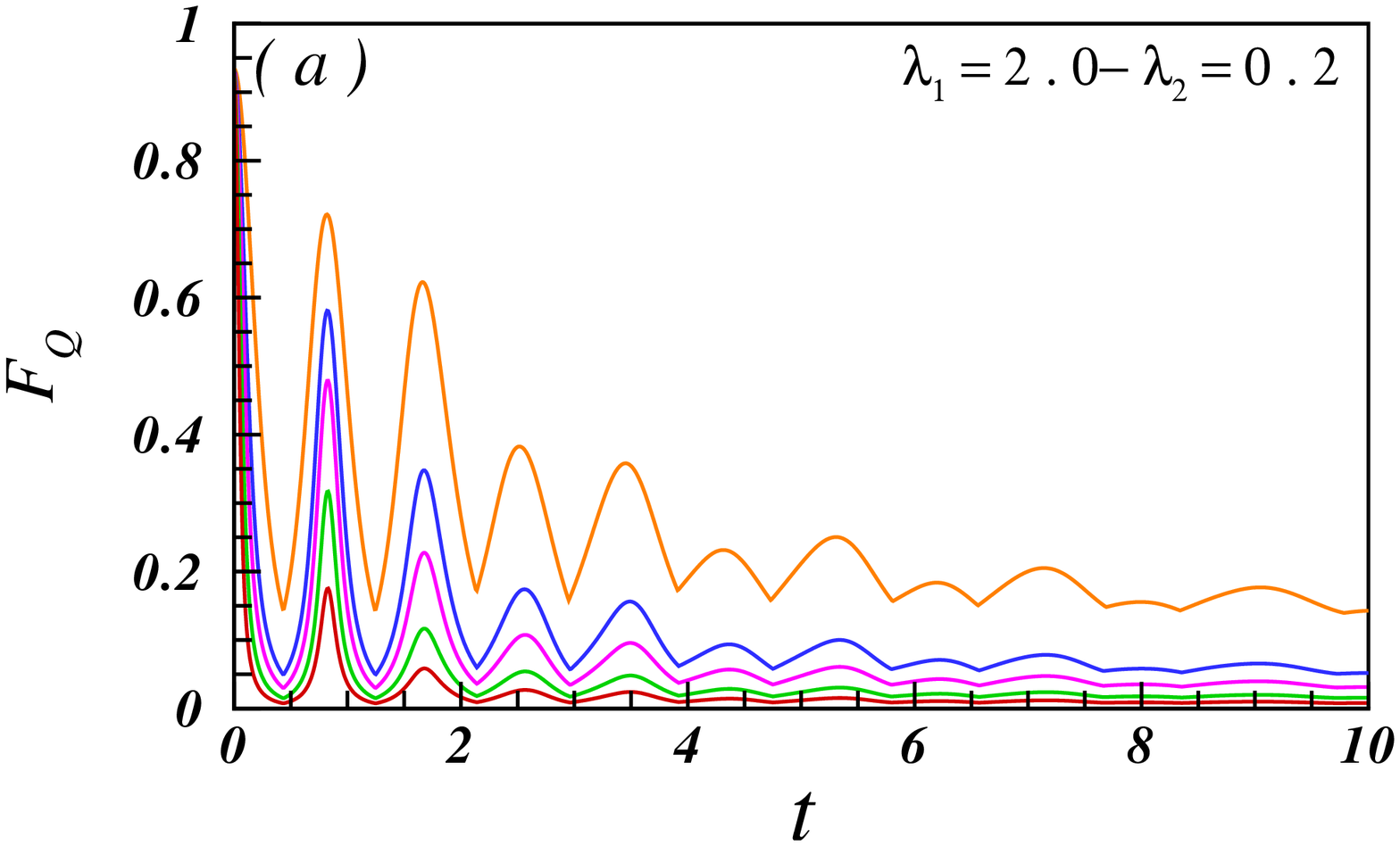}
\includegraphics[width=0.35\linewidth]{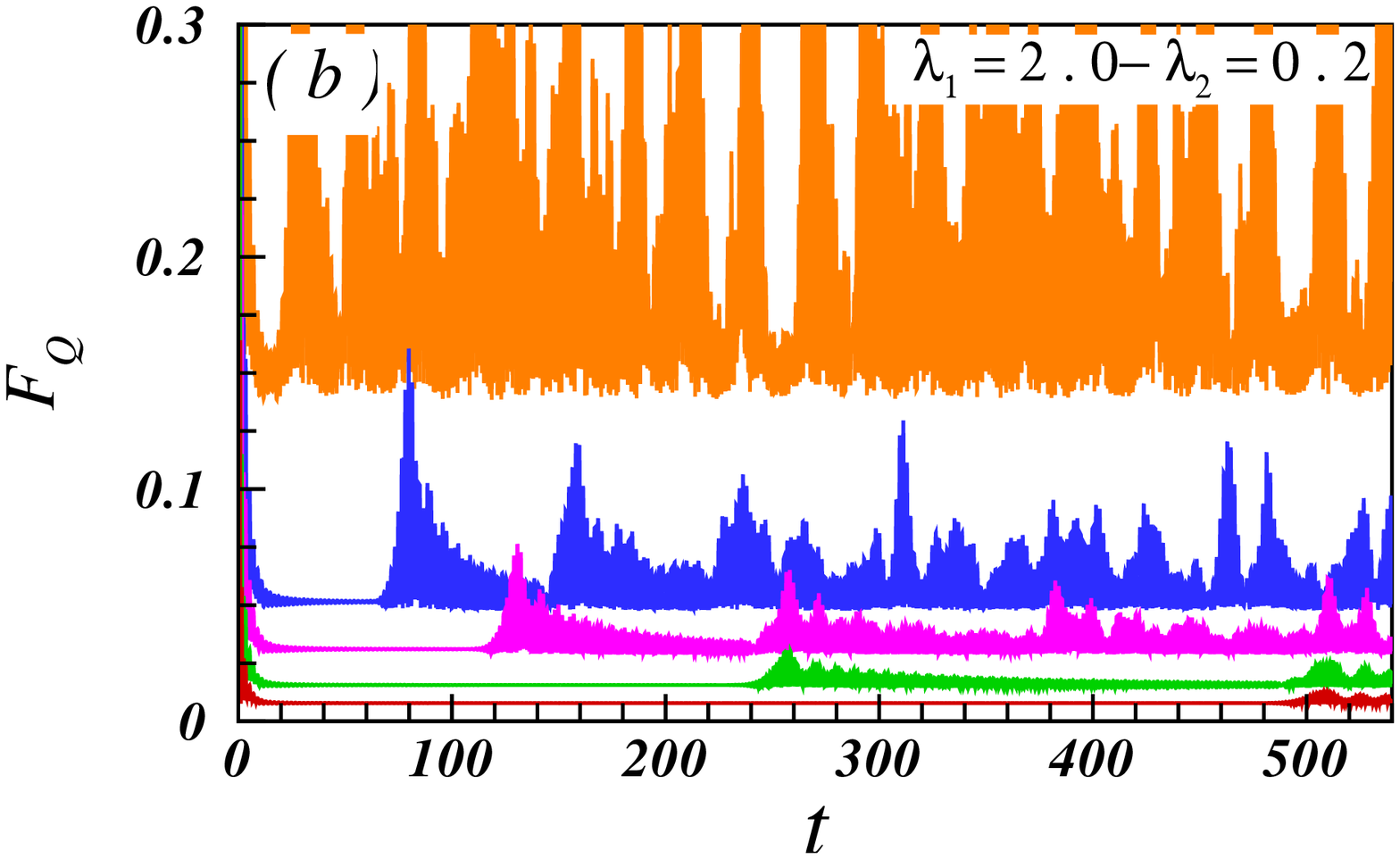}
\includegraphics[width=0.35\linewidth]{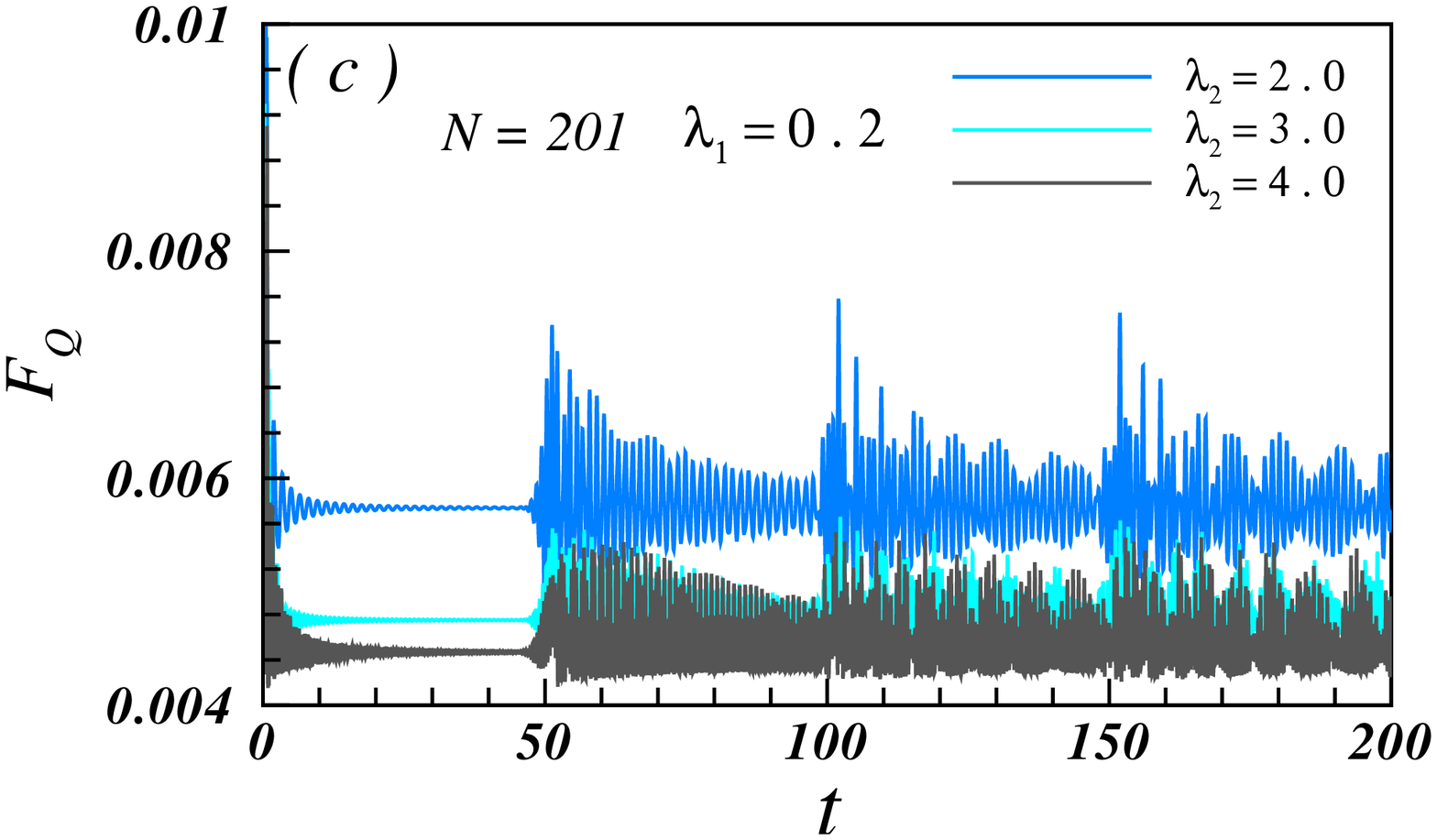}}
\caption{(color online). Dynamics of MQFI measured by ${\cal F}_{\cal Q}$. (a \& b) are drawn for quenches from $\lambda_1=2.0$ to $ \lambda_2=0.2$ for sizes $N=21,61,101,201,401$ (from orange to red).  (c) is for quenches from $\lambda_1=0.2$ to $ \lambda_2=2.0,3.0,4.0$ for the fixed size $N=201$.}
\label{Fig4}
\end{figure*}

Besides, at the first time it was shown in a dynamical quantum system envolving with a  quantum quench, the revival of a given state can be characterized by the survival probability of the state, as a consequence of a quench to a QCPs where the survival probabilities are related to the LE [{\color{blue}\onlinecite{{24}}}]. Later, it  was likewise demonstrated for other quenched points after that system reaches equilibrium state, the system again revivals and starts to fluctuate.
In general, the revival time for any arbitrary quench  is associated with the propagation of the state along the system, with a finite
$v_g(k)$ that is usually the velocity of the fastest quasiparticles [{\color{blue}\onlinecite{{55}}}], but this is not a rule and sometimes other quasiparticles can carry more information than the fastest quasiparticles such as what was discussed in the context of entanglement entropy [{\color{blue}\onlinecite{{56}}}], and LE [{\color{blue}\onlinecite{{57,58}}}]. The results give an approximate relationship for the revival time scale for the periodic condition as
\begin{eqnarray}\label{eq16}
T_{rev} \simeq \frac{N}{2v_g(k)}~,
\end{eqnarray}
For Hamiltonian ({\color{blue}\ref{eq1}}), it is easy to show $v_g(k)$ corresponds to the maximal group velocity $v_{max}=\mathop {\max }\limits_k  v_g(k)$ where ${v_g}(k) = \left| {\frac{{\partial {\varepsilon _k}}}{{\partial k}}} \right|$. Accordingly, $T_{rev}  = \frac{N}{2}\left| {\frac{\partial \varepsilon _k}{\partial k}} \right|_{max }^{ - 1}$.
It was proved the revival time for the LE is dependent on the final quenched point. Then, the initial point and hence the size of the quench i.e., the amplitude of the difference between pre-quench and post-quench control parameters $|\lambda_1-\lambda_2|$, are ineffective.
As a result, the revival  times are irrelevant to the critical times emerged from the DQPTs.
A simple calculation shows for the mentioned model in a final quenched point, $v_{max }=2 \lambda_2$ when  $\lambda_2  \le 1$ and $v_{max }=2 $  when  $\lambda_2  \ge 1$  that yield to $T_{rev}=\frac{N}{4\lambda_2}$ and $T_{rev}=\frac{N}{4}$ respectively. Remarkably, arising these revivals makes the quantum fidelity as a general probe to recognize the phase transitions and their universalities  [{\color{blue}\onlinecite{{59,60,61}}}]. 
 In the following, we seek the properties of the dynamical behavior of MQFI measured by  ${\cal F}_{\cal Q}(t)$ for the different system sizes under various quenches and compare obtained results with the results of the LE.


\subsection{Quench to the Critical Point}

Let us first consider the case of quenching from an arbitrary initial point $\lambda_1$ into the QCP $\lambda_2=\lambda_c=1$.  
Fig.~{\color{blue}\ref{Fig2}} presents quenches into the QCP  for different sizes as $N=21, 61, 101, 201, 401$. We put the system in (a) the FM phase, and  (b) the PM phase, with  $\lambda_1=1.5$ and $\lambda_1=0.5$ where ${\cal F}_{\cal Q}$ has remarkable  and small values respectively. 
As is clear, Fig.~{\color{blue} \ref{Fig2}} (a) shows for quenches started from initial states in $\lambda_1>\lambda_c$, the ${\cal F}_{\cal Q}$ exhibits oscillatory behavior consisting of a rapid decay followed by a revival.
The revivals emerge with a periodicity oscillation $T_{rev}$ exactly the same as what arises for the LE [{\color{blue}\onlinecite{{24}}}].  It turns out that as $N$ increases the values of revivals will decrease, while the revival time enhances.
On the other hand, for initial states chosen in $\lambda_1<\lambda_c$, Fig.~{\color{blue} \ref{Fig2}} (b) displays that ${\cal F}_{\cal Q}$ reveals a rapid enhance  followed by a decay at a  decay time as $T_{dec}$.
In this case, the behavior of the decay time is similar in the revival time, that is, by increasing the value of the system size, the decay time also increase but the value of ${\cal F}_{\cal Q}$ decrease. 
As a result  in the literature of macroscopic superposition, thermodynamics limit gives ${\cal F}_{\cal Q} \to 0$, that results in $p<2$. It means, only at a finite system size, there will be macroscopic superposition.
The interesting result is for the system with a certain size, the revival time matches with decay time, independent on the initial state and the size of the quench in such a way that at the revival time, where the initial state is located at the FM phase, ${\cal F}_{\cal Q}$ is minimum, while at the decay time, where the initial state is located at the FM phase, ${\cal F}_{\cal Q}$ is maximum. This means $T_{dec}=T_{rev}$.
This absolutely origins from the universality although the shape of the structure of quenching into the QCP is dependent on the
initial phase of equilibrium.  In addition, in Fig.~{\color{blue} \ref{Fig2}} (c), it is illustrated the scaling of the first revival and decay period times $T_{rev}$ and $T_{dec}$ respect to the system size that uncovers $T_{{rev}/{dec}}$  behaves almost linearly with $N$ as $T_{{rev}/{dec}}  \propto N$. 
As a consequence, this so-called critically enhanced decay and decreased revival of ${\cal F}_{\cal Q}$ for  quenching to the QCP can be suggested as a tool to probe of quantum phase transitions in the quantum many-body systems.

In order to show a more explicit  image of matching revival and  decay times in the dynamical behavior of LE and MQFI when a quench is done into the QCP, in Fig.~{\color{blue} \ref{Fig3}} (a \& b), we have displayed LE and ${\cal F}_{\cal Q}$ for quenches from (a) $ \lambda_1=1.1$, and (b) $ \lambda_1=0.9$ for size $N=201$. These two figures explicitly disclose  for a given size, these times are quite identical.
Moreover, as is driven from the results, the structures of revival and decay phenomena for quenches into the QCP obey from the universality. This is since the group velocity depends only on the quasiparticle dispersion, and other details such as the initial state and the size of the quenches are irrelevant.
We are stressed again, as explicitly seen from Fig.~{\color{blue} \ref{Fig3}} (a \& b), the shape of the structures in quenching into the QCP depends on the initial phase.
To more clarity,  in Fig.~{\color{blue} \ref{Fig3}} (c), as a pattern, we have plotted quenches from different initial states corresponding to different parameter values $\lambda_1=0.0, 0.5, 0.7, 0.9$  of the PM phase, into the QCP for size $N=201$. As one can see  the shape of the structures of the plots are the same for all the quenches,  and the periodic decay times are consistent with those predicted by  ({\color{blue}\ref{eq16}}). Moreover, for quenches near to the QCP, the value of ${\cal F}_{\cal Q}$  will be bigger.
 For the purpose of exhibiting the independent of the size of the quenches, in the insert in (c), we have  depicted the first decay time $T_{dec}$ versus $ |\lambda_2-\lambda_1|$ for quenching from region $0 \le {\lambda _1} < 1$ into $ \lambda_2=1.0$ for sizes $N=61,101,201,401$ (from red to cyan). This clearly indicates for any given system size, the value of the decay time remains constant  as $T_{dec}=\frac{N}{4}$, revealing independence of the decay times respect to the size of the quenches.
Our outcomes imprint  the dynamical behavior of ${\cal F}_{\cal Q}$  for quenching into the QCP is completely compatible with what is reported for the LE [{\color{blue}\onlinecite{{24,61}}}].


\subsection{Quench Crossed from the  Critical Point}
Quenching crossed from the equilibrium phase transition points makes an interesting choice of the quenching protocol.
The initial behavior of the MQFI is declining when a quench is done from the FM phase or from the PM phase into the FM phase. On the other hand, for quenching from the PM phase into the same phase, it is increasing.
In Fig.~{\color{blue} \ref{Fig4}} (a \& b), we have plotted the time evolution of MQFI  measured by  ${\cal F}_{\cal Q} $  for different sizes as $N=21,61,101,201,401$. The plots are for quenches from $\lambda_1=2.0$ to $ \lambda_2=0.2$.
As can be viewed,  (a) exposes for a given quench, as the system size enhances the ${\cal F}_{\cal Q} $  reduces quicker in a short time and afterward,  fluctuates slower. At the thermodynamics limit, by passing time, the fluctuations pretty much dissipate. 
It should be stressed the value of ${\cal F}_{\cal Q} $ in the decaying behavior tends to zero but not be exactly zero, even in the system at thermodynamic limit.
On the other side, Fig.~{\color{blue} \ref{Fig4}} (b) shows the structure of plots and consequently  revival times in the long time evolution complies ({\color{blue}\ref{eq16}}), and accepts the impact of both the system size and final quenched point. Besides, for quenching from the PM phase to the FM phase, enhancing the size decreases the value of ${\cal F}_{\cal Q} $, wipes out the oscillations, and makes the curve smooth. Anyway, in Fig.~{\color{blue} \ref{Fig4}} (c) it is illustrated quenching from $\lambda_1=0.2$ to $ \lambda_2=2.0,3.0,4.0$ for size $N=201$. In this case since  $\lambda_2  \ge 1$, thus for given system size, the final quenched point or in  other words  the size of the quench is ineffectiveness in dynamics of ${\cal F}_{\cal Q} $ because of $\lambda_2  \ge 1$ and hence $T_{rev}=\frac{N}{4}$. In addition,  Fig.~{\color{blue} \ref{Fig1}} demonstrates in the FM phase at the equilibrium state at $t=0$,  ${\cal F}_{\cal Q} $ is nearly independent of the system size when $\lambda  > \lambda_c$  [{\color{blue}\onlinecite{62}}]. In contrast, as driven from discussed results up to now, this statement is unacceptable when the system evolves with the passing time for all quenches, either from/into the FM phase or from/into the PM phase.
As a consequence from the perspective of macroscopic superposition, when a quench crosses  the QCP, at the thermodynamic limit, we have $p<2$, and hence the system cannot be macroscopic superpositions.

\begin{figure}[t]
\centerline{
\includegraphics[width=0.91\linewidth]{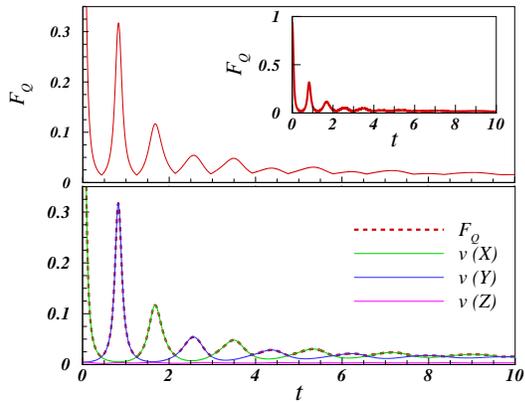}}
\caption{(color online). Dynamics of MQFI  measured by  ${\cal F}_{\cal Q}$ and  the variance ${{\cal V}_\Psi }(\eta (t)) $ with $\eta =X,Y,Z$. The plots are for $N=201$ and a quench from $ \lambda_1=2.0$ to  $ \lambda_2=0.2$. For this given quench, the cusps sit at the minimum values of ${\cal F}_{\cal Q}$   where the variances intersect.
}
\label{Fig5}
\end{figure}

\begin{figure}[t]
\centerline{
\includegraphics[width=0.87\linewidth]{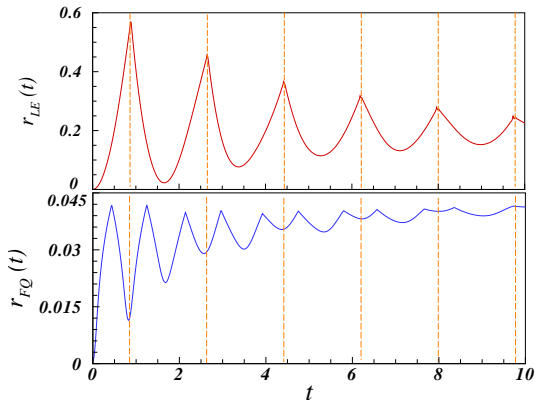}}
\caption{(color online).  The plots are for $N=201$ and a quench from $ \lambda_1=2.0$ to  $ \lambda_2=0.2$. The red is for $r_{LE}(t)$ while the blue corresponds to $r_{{\cal F}_{\cal Q}}(t)$. As shown, the first minimum of $r_{{\cal F}_{\cal Q}}(t)$  accurately happens at the first critical time $r_{LE}(t)$.
}
\label{Fig6}
\end{figure}

We have found in some quenches that are crossed from the QCP, in the real time evolution of MQFI, some non-analytical points can arise as cusps. This phenomenon occurs because of changing of the local observable from one direction to another to detect MQFI. We nominate this phenomenon  \textit{the dynamical MQFI transition} that happens at the critical times, $t_c$. 
As a sample, in Fig.~{\color{blue} \ref{Fig5}} we have displayed a quench from $ \lambda_1=2.0$ to  $ \lambda_2=0.2$ for size $N=201$.  The  insert in the top figure shows the dynamics of ${\cal F}_{\cal Q}$. In the top figure we have plotted ${\cal F}_{\cal Q}$ in the region $[0,0.35]$ to  show more clearly the existence of cusps where they stay on the mimimum values of ${\cal F}_{\cal Q}$ for this given quench. As explicit from the bottom figure, the cusps are formed due to the intersection between the variances of $\eta (t)$ which are defined as ${{\cal V}_\Psi }(\eta (t)) = \left\langle {{\eta ^2(t)}} \right\rangle  - {\left\langle {\eta (t)} \right\rangle ^2}$ with $\eta =X,Y,Z$. In general, the critical times $t_c$ are not periodic and any number of them may appear at different critical times.

\begin{figure}[t]
\centerline{
\includegraphics[width=0.902\linewidth]{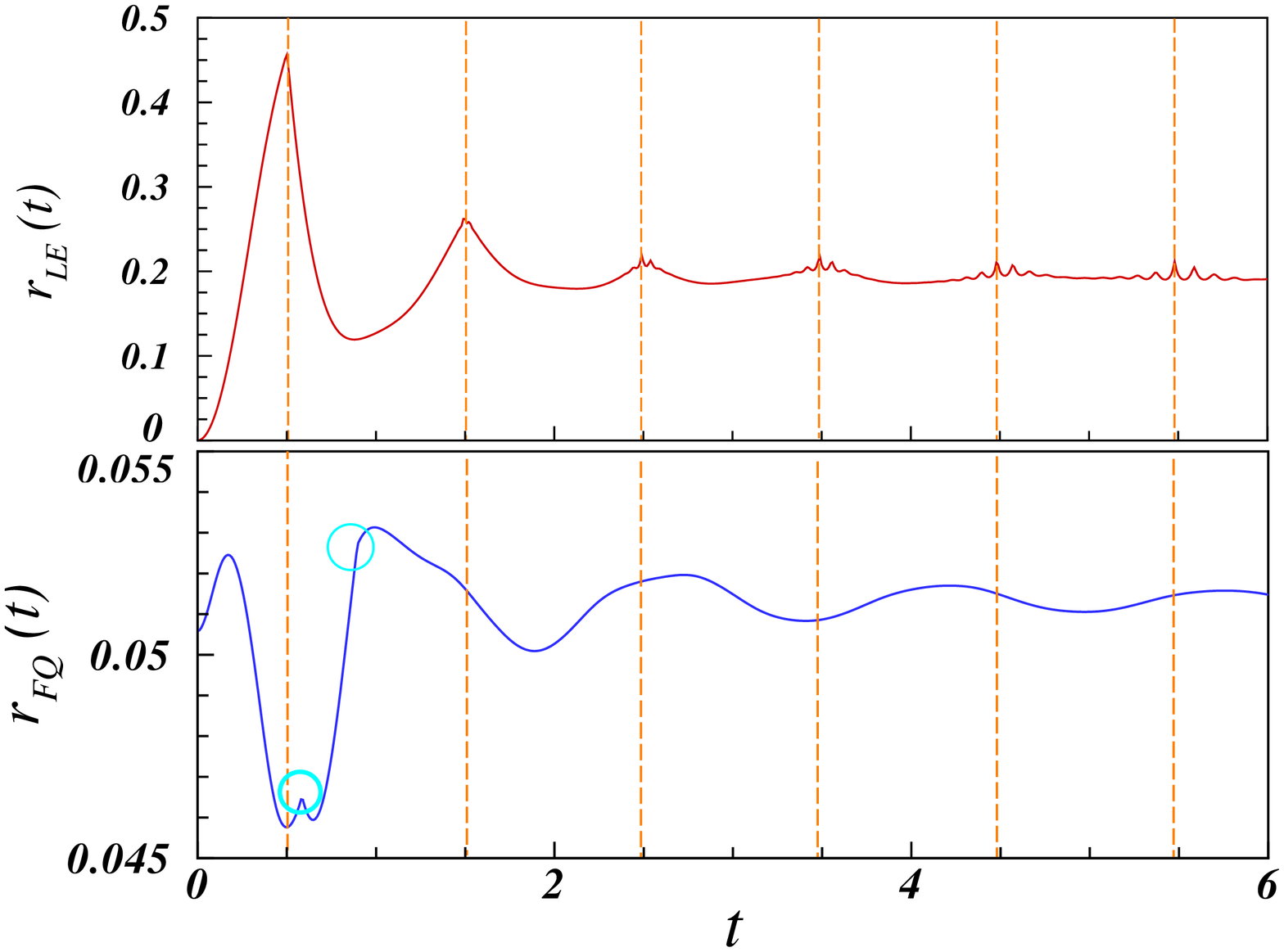}}
\caption{(color online). The plots are for $N=201$ and a quench from $ \lambda_1=0.2$ to  $ \lambda_2=2.0$. The red is for $r_{LE}(t)$ while the blue corresponds to $r_{{\cal F}_{\cal Q}}(t)$. As seen, the first minimum of $r_{{\cal F}_{\cal Q}}(t)$  accurately occurs at  the first critical time $r_{LE}(t)$. The cyan circles hint the  cusps at the critical times $t_c$.  }
\label{Fig7}
\end{figure}

\begin{figure*}[t]
\centerline{
\includegraphics[width=0.374\linewidth]{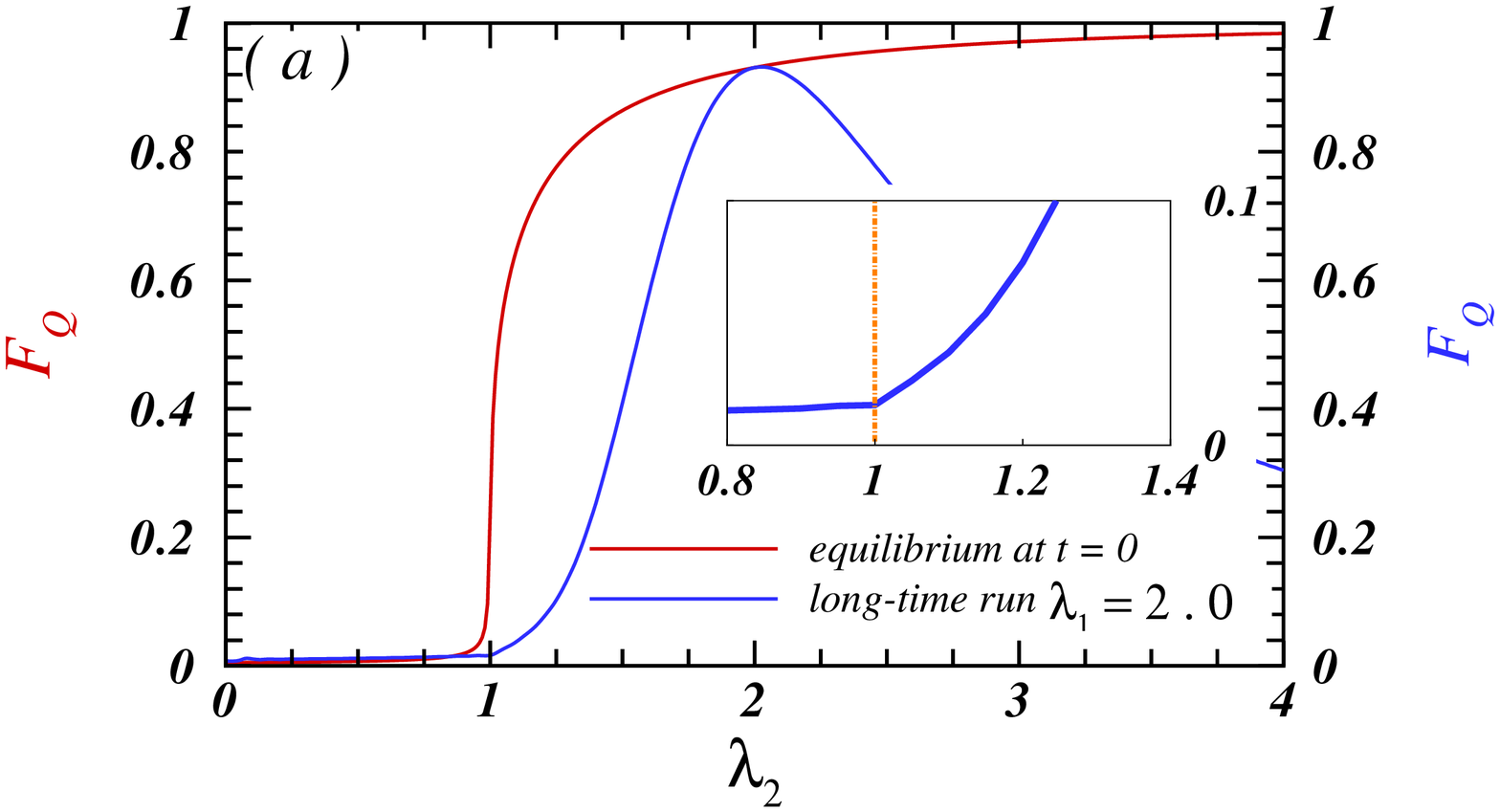}
\includegraphics[width=0.376\linewidth]{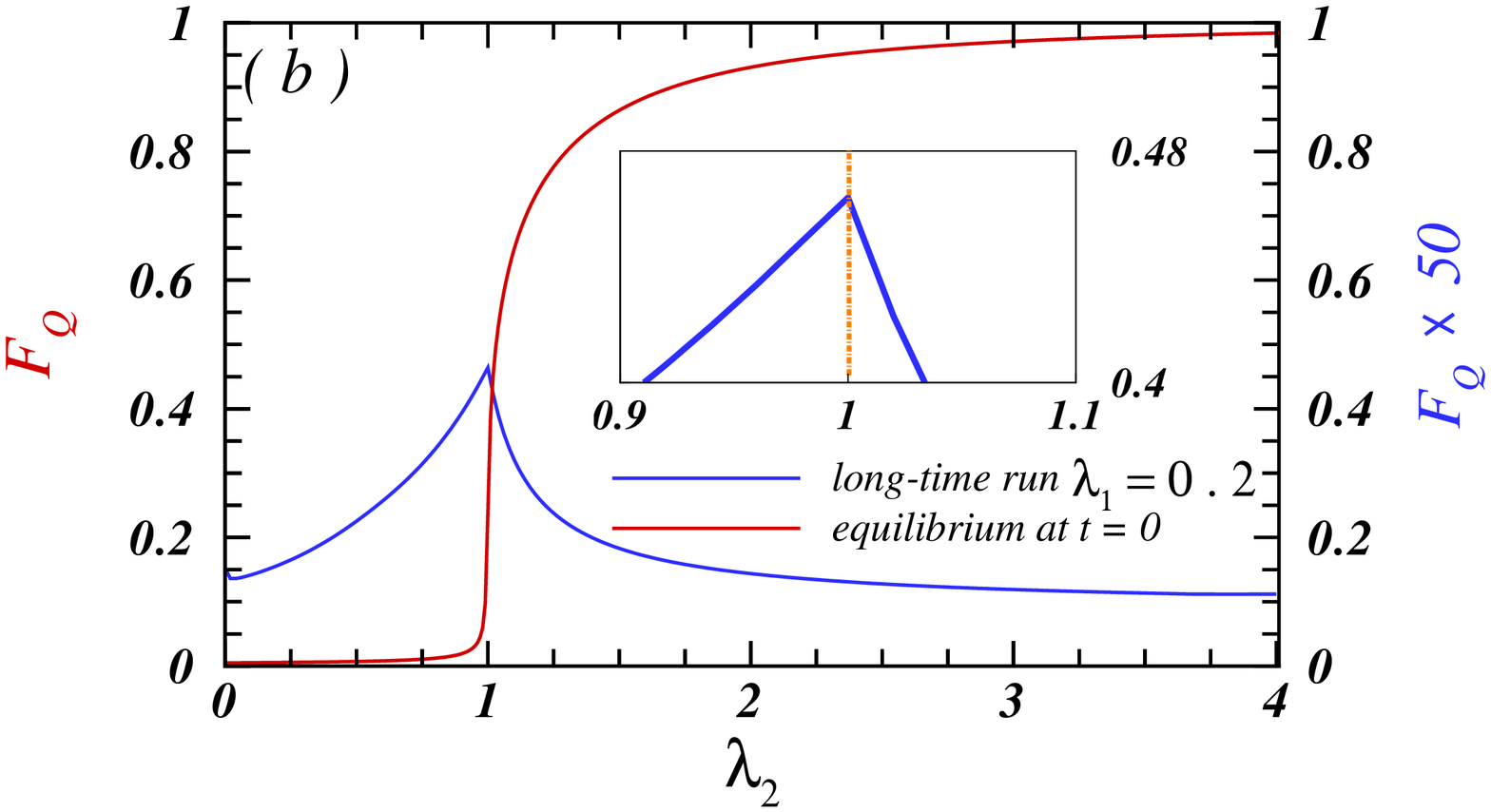}}
\caption{(color online). The equilibrium situation at $t=0$ (the red) and long-time run  (the blue) of   ${\cal F}_{\cal Q}$ as function of $ \lambda_2$ for size $N=401$ for quenches from (a) $ \lambda_1=2.0$, and (b) $ \lambda_1=0.2$  respectively. The plot in (b) for long-time run scales fifty times larger. The inserts exactly show the signatures  of a nonequilibrium quantum phase transition  at  $ \lambda_c=1.0$.
}
\label{Fig8}
\end{figure*}

It can be drived from Figs.~{\color{blue} \ref{Fig2}} \& {\color{blue}\ref{Fig4}} where for $t>0$ the value of ${\cal F}_{\cal Q} $ is non-zero at all times. Thus, there is no Fisher zero in the dynamics of the MQFI. To be more accurate, similar in DQPT, we define  a logarithmic function of ${\cal F}_{\cal Q} $ in the form
\begin{eqnarray}\label{eq17}
r_{{\cal F}_{\cal Q}}(t) =  - \frac{1}{N}\log \left| {{\cal F}_{\cal Q}(t)} \right|^2. 
\end{eqnarray}
In Figs.~{\color{blue} \ref{Fig6}} \& {\color{blue}\ref{Fig7}} we have depicted $r_{{\cal F}_{\cal Q}}(t)$ and $r_{LE}(t)$ for size $N=201$ for quenching from (Fig.~{\color{blue} \ref{Fig6}}) $ \lambda_1=2.0$ to  $ \lambda_2=0.2$, and (Fig.~{\color{blue}\ref{Fig7}})  $ \lambda_1=0.2$ to  $ \lambda_2=2.0$.  The cusps explicitly are visible in the time evolution of $r_{LE}(t)$.
As  mentioned there some cusps may  reveal in ${\cal F}_{\cal Q}$ at the critical times $t_c$ of the dynamical MQFI transition when  a quench crosses  the QCP.  The existence of the  non-analytical points also are visible exactly at the critical times $t_c$ in $r_{{\cal F}_{\cal Q}}(t)$ (please compare ${\cal F}_{\cal Q}$ and $r_{{\cal F}_{\cal Q}}(t)$ in Figs.~{\color{blue} \ref{Fig5}} \& {\color{blue}\ref{Fig6}}).
 Anyway, Fig.~{\color{blue}\ref{Fig7}} indicates two cusps at  $t_c=0.58, 0.89 $ (marked with the cyan circles).
Consequently, Figs.~{\color{blue} \ref{Fig6}} \& {\color{blue}\ref{Fig7}}  unveil the critical times coming out of the DQPT are difference with  the critical times emerging of the logarithmic function of ${\cal F}_{\cal Q}$.  That is, $t_c  \ne t^*_n$.
Hence, special mode $k^*$ that leads to vanishing the LE is unable to vanish ${\cal F}_{\cal Q}$  and thus there is no  Fisher zero.
Moreover, as clear, because of choosing the initial states in different phases, quenching from the FM phase to the PM phase (Fig.~{\color{blue} \ref{Fig6}}) makes different behaviors for $r_{{\cal F}_{\cal Q}}(t)$ rather than when a quench is performed in the opposite direction (Fig.~{\color{blue} \ref{Fig7}}).
However, the intriguing result is the first critical time arisen from $r_{LE}(t)$ is exactly equal to the first time whose $r_{{\cal F}_{\cal Q}}(t)$ is  minimum.
In addition, As viewed from Figs.~{\color{blue} \ref{Fig6}} \& {\color{blue}\ref{Fig7}}, for the times greater than the mentioned first extremums,  the other extremum times unfolded from $r_{{\cal F}_{\cal Q}}(t)$ are not necessarily the same as the critical times emerged from $r_{LE}(t)$.


\subsection{Long-time Run Behavior}

As mentioned the DQPTs are characterized by the emergence of Fisher zeroes at critical times during time evolution. Usually, the time average of the order parameter is used to characterize nonequilibrium criticality [{\color{blue}\onlinecite{{63}}}]. Additionally, it is  demonstrated that nonequilibrium quantum phase transitions can be identified by nonanalyticities in the long-time average of the LE [{\color{blue}\onlinecite{{64}}}]. These nonanalytic behaviors are illustrated by a sharp change.
To this end, for searching this feature in the dynamical behavior of MQFI in Fig. {\color{blue}\ref{Fig8}} we have considered the long-time run of ${\cal F}_{\cal Q} $ in the quench process described by the sudden change from the order and disorder phases with $ \lambda_1=2.0$ and $ \lambda_1=0.2$ to desired $ \lambda_2$  of  the final Hamiltonian for $N=401$ respectively. 
 The long-time run, ${\cal T}_{ltr}$, is the time that the dynamics of MQFI goes to or  fluctuates quietly around a stable situation.
Fig. {\color{blue}\ref{Fig8}} (a) corresponding to $ \lambda_1=2.0$ and ${\cal T}_{ltr}=20$ shows for $ \lambda_2<\lambda_c$ the long-time run value is almost equal to its equilibrium value at $t=0$, both are nearly to zero. As soon as the value of $ \lambda_2$ increases more than $ \lambda_c$, the  long-time run value  enhances, marking the quantum phase transition. The inset in (a) displays it clearly. Anyway, at  $ \lambda_2 \to \infty $ it decreases and goes to zero value. In other words, since the long-time run value  in region $ \lambda_2=(\lambda_c, \lambda>\lambda_1)$ is remarkable  for a finite value of  $ \lambda$, the macroscopic superposition states will emerge at the long-time run when a quench is done from the order phase within itself. In this case, the $p\textendash$index will be 2. This is the only situation in the quench dynamics of the system that the macroscopic states can be come into being.
On the other side, Fig. {\color{blue}\ref{Fig8}} (b) exhibits  the system with the initial state sitting in $\lambda_1=0.2$ at ${\cal T}_{ltr}=80$. Here, because the value of long-time run of ${\cal F}_{\cal Q} $ is small, for a clear presentation of the results, we scale its value fifty times larger. The plot shows that the long-time run value of ${\cal F}_{\cal Q} $ has an obvious change around the transition point. In the inset in (b), it is evident.
Meanwhile, as viewed, for $ \lambda_2<\lambda_c$, the ${\cal F}_{\cal Q} $ increments  as $ \lambda_2$ boosts, at $ \lambda_c$ will be maximum, and afterward for $ \lambda_2>\lambda_c$ abruptly declines. As a result, for a quench started from an initial state in the disordered phase since its long-time run is small, hence, the system is unable to have macroscopic superpositions.
Consequently, the sharp change of long-time run of MQFI at the transition point can give us a characteristic signature of the nonequilibrium quantum phase transition.


\section{Summary}

Quantum coherence can be regarded as a fundamental mark of non-classicality in physical systems.
It provides a strong framework for studying properties of quantum systems and their applications for quantum technology, especially in optical lattices. Despite several works on quantum coherence, the dynamics of the functions whose measure it has not yet been studied sufficiently.

In this paper, we have considered the dynamics of quantum coherence after sudden quantum quenches by employing MQFI and LE, and  probed what relationship would exist between these two functions.
We have found appealing results as the existence of a  relationship between them that will certainly lead to a better comprehension of the concept of coherence, especially in the dynamics of the non-equilibrium systems.
The same as the LE, a nontrivial revival structure will be emerged in the dynamical behavior of MQFI  that cannot be obtained from a simple spectral analysis. These structures adhere to the promised universality i.e., the initial state and the size of the quench are irrelevant, and relate to quasiparticles propagating of the system with a  maximum speed $v_{max}$, and also the system size [{\color{blue}\onlinecite{{55}}}]. Intriguingly, the critically enhanced and decreased of MQFI at revival and decay times are appeared by quenching from any arbitrary phase point of the FM and PM phases into the quantum phase transition respectively. We offer these features as a tool to discover  QCPs [{\color{blue}\onlinecite{{24}}}].
Anyway,  for quenching into the QCP for a system with given size, the universality leads to $T_{rev}=T_{dec}$. 
The interesting outcome corresponds to the quenches crossed from the QCP in such a way that for some of them, nonanalytic behaviors at critical times show up that we name them \textit{the dynamical MQFI transition}. 
Additionally, to examine more precisely the existence of Fisher zeroes in the time evolution of MQFI, we have established a logarithmic  function of MQFI. Our analyses have indicated although no Fisher zeroes appear when a quench  crosses  the QCP, the first time whose the logarithm of MQFI is minimum is exactly equal to the first critical time that emerged from the  DQPT [{\color{blue}\onlinecite{{35}}}].
Moreover, we have figured out the only situation in the quench dynamics of the system that the macroscopic superposition states can be arisen is quenching from the FM phase within the same phase, but not for very far away from the initial state [{\color{blue}\onlinecite{{40}}}].
We further have imprinted the long-time run behavior of MQFI is able to reveal the nonequilibrium quantum phase transition [{\color{blue}\onlinecite{{64}}}].
We hope the results help to a deeper understanding of the coherence dynamics and give a more clear image of quantum coherence measurement tools in the quantum systems far away from equilibrium.

\vspace{0.3cm}


\end{document}